\scrollmode

\documentclass[aps,prl,groupedaddress,superscriptaddress,amsmath,amssymb,showpacs,twocolumn,stfloats]{revtex4}
\usepackage{graphicx}
\usepackage{bm}
\usepackage{color}
\usepackage{times,txfonts}
\usepackage{hyperref}

\def\ketbra#1#2{\vert #1\rangle\langle #2\vert}
\def\ket#1{\vert #1\rangle}

\def\C{\mathbb{C}}

\def\cH{\mathcal{H}}

\begin{document}
\title{Strong monogamy conjecture for multiqubit entanglement: The four-qubit case}
\author{Bartosz Regula}
\affiliation{School of Mathematical Sciences, The University of Nottingham, University Park, Nottingham NG7 2RD, United Kingdom}
\author{Sara Di Martino}
\affiliation{Dipartimento di Matematica, Universit\`{a} di Bari, I-70125 Bari, Italy}
\affiliation{School of Mathematical Sciences, The University of Nottingham, University Park, Nottingham NG7 2RD, United Kingdom}
\author{Soojoon Lee}
\affiliation{Department of Mathematics and Research Institute for Basic Sciences, Kyung Hee University, Seoul 130-701, Korea}
\affiliation{School of Mathematical Sciences, The University of Nottingham, University Park, Nottingham NG7 2RD, United Kingdom}
\author{Gerardo Adesso}
\affiliation{School of Mathematical Sciences, The University of Nottingham, University Park, Nottingham NG7 2RD, United Kingdom}
\begin{abstract}
We investigate the distribution  of bipartite and multipartite  entanglement in multiqubit states. In particular we define a set of monogamy inequalities sharpening the conventional Coffman--Kundu--Wootters constraints, and we provide analytical proofs of their validity for  relevant classes of states. We present extensive numerical evidence validating the conjectured strong monogamy inequalities for arbitrary pure states of four qubits.
\end{abstract}

\date{\today}
\maketitle

\begin{figure}[t]
  \centering
    \includegraphics[width=7.5cm]{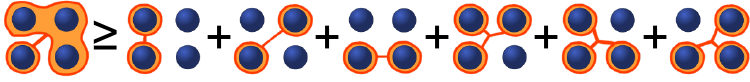} \\[-0.3cm]
 \caption{
 Strong monogamy of four-qubit entanglement.
 \label{fig:monineq}\vspace*{-.4cm}}
\end{figure}

{\it Introduction.}---
Entanglement is one of the most fundamental and intriguing features of quantum mechanics. It plays a crucial role for applications in  quantum information and communication, and in a variety of  areas ranging from quantum field theory to condensed matter, statistical physics, thermodynamics, and biology \cite{review_horodecki,review,jensreview}. Despite its central importance, however, the physical understanding and mathematical description of its essential characteristics remain highly nontrivial tasks, particularly when many-particle systems are analyzed.

One of the key properties distinguishing entanglement from classical correlations is its {\it monogamy}: entanglement cannot be freely shared among multiple parties \cite{barbara}. Monogamy is a consequence of the  no-cloning theorem \cite{nocloning,gamonorev,clonprl}, and is  obeyed by several types of nonclassical correlations, including Bell nonlocality \cite{toner}, Einstein--Podolsky--Rosen steering \cite{monosteer} and contextuality \cite{monocontext}, albeit not discord-type correlations \cite{arequantum}.

In 2000 Coffman, Kundu, and Wootters (CKW) formalized the monogamy of entanglement for a three-qubit system in the form of a quantitative constraint, known as `monogamy inequality' \cite{monogamy}. Given an arbitrary pure state $\ket{\psi}\in\cH=(\C^2)^{\otimes 3}$ of three qubits $q_1$, $q_2$, and $q_3$, the CKW inequality with respect to the choice of $q_1$ as a focus can be written as
\begin{equation}
\label{eq:ckw_ineq}
\tau^{(1)}_{q_1|(q_2q_3)}(\ket{\psi}) \geq \tau^{(2)}_{q_1|q_2}(\ket{\psi})+\tau^{(2)}_{q_1|q_3}(\ket{\psi}).
\end{equation}
Here $\tau^{(2)}_{q_i | q_j}$ denotes the bipartite entanglement in the reduced state of the pair of qubits  $q_i$ and $q_j$, quantified by a computable entanglement monotone known as two-tangle, or simply tangle \cite{concurrence, concurrence2,monogamy}. The term  $\tau^{(1)}_{q_1|(q_2q_3)}$ denotes the one-tangle, a measure of entanglement between  $q_1$ and the rest of the system, given by the linear entropy of the marginal state of qubit $q_1$,
\begin{equation}\label{eq:onetangle}
\tau^{(1)}_{q_1|(q_2 q_3)}(\ket{\psi})= 4 \det \rho_1,
\end{equation}
where $\rho_{1}=\mathrm{Tr}_{q_2 q_3}(\ketbra{\psi}{\psi})$ is the density matrix of qubit $q_1$ and $\mathrm{Tr}_X(\rho)$ indicates the partial trace of $\rho$ over subsystem $X$.

The meaning of Eq.~(\ref{eq:ckw_ineq}) is clear: the entanglement between $q_1$ and the two other qubits taken as a group cannot be less than the sum of the individual entanglements between $q_1$ and each of the two remaining qubits. Similar inequalities can be written by selecting $q_2$ or $q_3$ as focus qubits. Remarkably, the difference between left and right hand side of (\ref{eq:ckw_ineq}) can be interpreted as a quantifier of the entanglement genuinely shared among the three qubits. Precisely, one can define the residual three-qubit tangle---or, in short, three-tangle---of $\ket{\psi}$ as
\begin{equation}\label{eq:res3}
\tau^{(3)}_{q_1|q_2|q_3}(\ket{\psi}):=\tau^{(1)}_{q_1|(q_2 q_3)}(\ket{\psi})-\tau^{(2)}_{q_1|q_2}(\ket{\psi})-\tau^{(2)}_{q_1|q_3}(\ket{\psi}).
\end{equation}
Interestingly, this quantity does not depend on the focus qubit (e.g., $q_1$) that we privilege in the decomposition. Namely, $\tau^{(3)}_{q_1|q_2|q_3}(\ket{\psi})=\tau^{(1)}_{q_2|(q_3 q_1)}(\ket{\psi})-\tau^{(2)}_{q_2|q_3}(\ket{\psi})-\tau^{(2)}_{q_2|q_1}(\ket{\psi})=\tau^{(1)}_{q_3|(q_1 q_2)}(\ket{\psi}) -\tau^{(2)}_{q_3|q_1}(\ket{\psi})-\tau^{(2)}_{q_3|q_2}(\ket{\psi})$ as well \cite{monogamy}. The three-tangle is a full-fledged measure of the genuine tripartite entanglement of any three-qubit pure state $\ket{\psi}$ \cite{dur,review_horodecki}.

A generalization of the CKW inequality (\ref{eq:ckw_ineq}) to $n$-qubit systems was only proven by Osborne and Verstraete \cite{genmon} several years after the original conjecture \cite{monogamy}. Denoting now by $\ket{\psi}$ a general pure state of $n$ qubits,
the following holds \cite{genmon},
\begin{equation}
\label{eq:gen_ckw}
\tau^{(1)}_{q_1|(q_2\cdots q_n)}(\ket{\psi}) \geq \tau^{(2)}_{q_1|q_2}(\ket{\psi})+\tau^{(2)}_{q_1|q_3}(\ket{\psi})+\ldots+\tau^{(2)}_{q_1|q_n}(\ket{\psi}).
\end{equation}
This means that the entanglement between $q_1$ and the rest  is not less than the sum of the individual pairwise entanglements involving $q_1$ and each of the other $n-1$ qubits $q_j$ ($j=2,\ldots,n$). However, for $n>3$, the difference between left and right hand side in (\ref{eq:gen_ckw}) just gives a rough indicator of all the leftover entanglement not distributed in pairwise form.
Attempts to construct generalized monogamy inequalities  in $n$-qubit systems have been considered \cite{gmchris,gmgour,gmcorn,jensreview}, but these have not led to clear recipes to isolate the genuine $n$-partite entanglement, nor have resulted in a general sharpening of (\ref{eq:gen_ckw}) for arbitrary states.

In this Letter we
propose and investigate a set of sharper monogamy constraints. We
raise the intuitive hypothesis that the residual in (\ref{eq:gen_ckw}) is amenable to a further decomposition into individual $m$-partite contributions which involve $m=3,4,\ldots,n-1$ qubits, in all possible combinations encompassing the focus qubit $q_1$. Heuristically, one can expect that all of these multipartite contributions be independent, overall adding up to the global bipartite entanglement between $q_1$ and the rest of the system.
This leads us to postulate a hierarchy of {\it strong monogamy} (SM) inequalities limiting the distribution of bipartite and multipartite entanglement in $n$-qubit systems, which take in general the following form
\begin{eqnarray}
\label{eq:strongmono}
&&\hspace*{-.4cm}\tau^{(1)}_{q_1|(q_2\cdots q_n)}(\ket{\psi})  \geq  \sum_{m=2}^{n-1} \sum_{\vec{j}^m}[\tau^{(m)}_{q_1|q_{j^m_1}|\cdots |q_{j^m_{m-1}}}(\ket{\psi})]^{\mu_{m}} \equiv \underbrace{\sum_{j=2}^n \tau^{(2)}_{q_1|q_j}(\ket{\psi})}_{\mbox{$2$-partite}} \nonumber \\
&&\hspace*{-.4cm}+ \underbrace{\sum_{k>j=2}^n [\tau^{(3)}_{q_1|q_j|q_k}(\ket{\psi})]^{\mu_3}}_{\mbox{$3$-partite}} + \ldots + \underbrace{\sum_{l=2}^n [\tau^{(n-1)}_{q_1|q_2|\cdots |q_{l-1}|q_{l+1}|\cdots|q_n}(\ket{\psi})]^{\mu_{n-1}}}_{\mbox{$(n-1)$-partite}},  \nonumber \\
\end{eqnarray}
 where we have employed a short-hand notation, introducing the index vector  $\vec{j}^m=(j^m_1,\ldots,j^m_{m-1})$ which spans all the ordered subsets of the index set $\{2,\ldots,n\}$ with $(m-1)$ distinct elements, and we have included in general a sequence of rational exponents $\{\mu_m\}_{m=2}^{n-1}$, with $\mu_2 \equiv 1$, which can regulate the weight assigned to the different $m$-partite contributions.

  Our main conjecture is that inequality (\ref{eq:strongmono}), and its variants for different choices of the focus qubit, hold simultaneously for arbitrary pure states $\ket{\psi}$ of $n$ qubits, provided one adopts a suitable definition of the  $m$-partite quantities $\{{\tau}^{(m)}, \mu_m\}$.  We remark that, for a given choice of the involved entanglement monotones (tangles), the expression in (\ref{eq:strongmono}) yields a whole class of monogamy constraints, parameterized by the powers $\mu_m$.  Any nontrivial selection of the sequence $\{\mu_m\}_{m=2}^{n-1}$ with $\mu_2 \equiv 1$ defines in fact a particular SM inequality,  sharpening and generalizing the CKW one. Clearly,  the verification of  (\ref{eq:strongmono}) given a set   $\{\mu_m^\star\}$ implies its validity for all $\{\mu_m\} \succeq \{\mu_m^\star\}$. For this reason, in order to establish the sharpest instance, one should aim to prove the inequalities by fixing each $\mu_m$ to be as small as possible, with $\mu_m=1$ $\forall m$ being the minimal choice.   We will specify the adopted choices of the parameters $\mu_m$ in the subsequent analysis.

Interestingly, a constraint alike to (\ref{eq:strongmono}) was shown to hold for the distribution of entanglement in permutationally-invariant continuous variable Gaussian states, leading to an operational quantification of genuine $n$-partite entanglement \cite{strongmonogauss}.
This  gives a strong hint that a similar sharing structure should hold for entanglement in finite-dimensional systems too, although no supporting evidence was obtained prior to this work.




{\it Setting up the notation.}--- Here we adopt the following prescriptions.
First, we define the pure-state residual $n$-tangle ${\tau}^{(n)}$  as the difference between left and right hand side in (\ref{eq:strongmono}),
 \begin{equation}\label{eq:taunpure}
 \tau^{(n)}_{q_1|q_2|\cdots|q_n}(\ket{\psi}):=\tau^{(1)}_{q_1|(q_2\cdots q_n)}(\ket{\psi}) - \! \sum_{m=2}^{n-1}\! \sum_{\vec{j}^m}[\tau^{(m)}_{q_1|q_{j^m_1}|\cdots |q_{j^m_{m-1}}}\!\!(\ket{\psi})]^{\mu_m}\,.
 \end{equation}
  In this way, the conjectured SM inequality (\ref{eq:strongmono}) is recast into the nonnegativity of the residual, ${\cal \tau}^{(n)}_{q_1|q_2|\cdots|q_n}(\ket{\psi}) \geq 0$, where the ordering of the subscripts in (\ref{eq:taunpure}) reflects the choice of the focus qubit, which occupies the first slot (we do not expect permutation invariance for $n>3$).  Next, we extend the residual $n$-tangle ${\cal \tau}^{(n)}$ to a mixed state $\rho$ of $n$ qubits via a  convenient and physically motivated convex roof procedure,
\begin{equation}\label{eq:taunmix}
{\cal \tau}^{(n)}_{q_1|q_2|\cdots|q_n}(\rho)
:=\bigg[\inf_{\{p_r, \ket{\psi_r}\}} \sum_r p_r \sqrt{\tau^{(n)}_{q_1|q_2|\cdots|q_n}(\psi_r)}\bigg]^{2},
 \end{equation}
 where the minimization is taken over all possible pure-state decompositions of the state $\rho=\sum_r p_r \ketbra{\psi_r}{\psi_r}$.
  For $n=3$, the definition (\ref{eq:taunmix}) reduces to the mixed-state extension of the three-tangle $\tau^{(3)}$ as defined in \cite{verstraete4}, which is an entanglement monotone \cite{monogamy,dur,review_horodecki} and an invariant under stochastic local operations and classical communication (SLOCC) \cite{verstraete4,verstraetebart,jens,jensreview}. For $n=2$, we recover the standard pairwise tangle,  $\tau^{(2)}_{q_i|q_j} = C^2_{q_i|q_j}$, with the concurrence  \cite{concurrence,concurrence2} defined as $C_{q_i|q_j}=\max\{0,\lambda_1-\lambda_2-\lambda_3-\lambda_4\}$,
where $\{\lambda_j\}$ are the square-roots of the eigenvalues (in decreasing order) of the matrix $R=\rho_{ij}(\sigma_y\otimes\sigma_y)\rho_{ij}^*(\sigma_y\otimes\sigma_y)$,
  the star denoting complex conjugation in the computational basis, $\sigma_y$ being the Pauli $y$ matrix, and $\rho_{ij}$ being the marginal state of qubits $q_i$ and $q_j$ obtained by partial tracing over the remaining qubits.
Finally, we use Eqs.~(\ref{eq:taunpure})--(\ref{eq:taunmix}) to define, in a recursive way, every $m$-partite term ${\tau}^{(m)}$ (for $m \geq 2$) appearing in the $n$-qubit SM inequality (\ref{eq:strongmono}), in terms of the corresponding residual $m$-tangle rescaled by a suitable exponent $\mu_m$.

Proving the SM conjecture for $n$ qubits appears in general a formidable challenge. Namely, at variance with the CKW case, the $m$-tangles defined above are not expected to enjoy a closed formula on the marginal $m$-qubit mixed states for $m \geq 3$. Nonetheless, in the following we verify the conjecture analytically on relevant multiqubit states, and we achieve significant progress on arbitrary states of four qubits ($n=4$), for which we provide a comprehensive collection of analytical and numerical evidence in support of the SM hypothesis.


{\it Analytical example: GHZ/W superpositions.}---
We begin by investigating the SM constraint (\ref{eq:strongmono}) in its sharpest form ($\mu_m=1$ $\forall m$), on permutationally invariant  states defined as superpositions of $W$ and generalized Greenberger-Horne-Zeilinger (GHZ) states of $n \geq 4$ qubits,
\begin{equation}\label{eq:gengg}
\ket{\Phi^n_{\alpha,\beta,\gamma}} := \alpha \ket{0^n} + \beta \ket{W_n} + \gamma \ket{1^n}\,,
\end{equation}
with $\alpha,\beta,\gamma\in \C$, $|\alpha|^2+|\beta|^2+|\gamma|^2=1$, where $\ket{W_n}=\frac{1}{\sqrt{n}}(\ket{0^{n-1}1}+\ldots+\ket{10^{n-1}})$ is the $n$-qubit $W$ state, $\big|{\Phi^n_{{1}/{\sqrt2},0,{1}/{\sqrt2}}}\big\rangle$ is the $n$-qubit GHZ states,
and $x^n$ denotes the string with $n$ equal symbols $x$.
Noting that we can rewrite the states as $\ket{\Phi^n_{\alpha,\beta,\gamma}} = \ket{0^{n-m}}\left(\alpha \ket{0^m} + \sqrt{\frac{m}{n}}\beta \ket{W_m}\right) + \sqrt{\frac{n-m}{n}} \beta \ket{W_{n-m}}\ket{0^m} + \gamma \ket{1^{n-m}}\ket{1^m}$, for $1 \leq m \leq n-1$, where $\ket{W_1} \equiv \ket{1}$, and observing in particular that for $\gamma=0$ all residual multipartite terms vanish, ${\tau}^{(1)}(\ket{\Phi^n_{\alpha,\beta,0}}) = (n-1) {\tau}^{(2)}(\ket{\Phi^n_{\alpha,\beta,0}})$, we obtain the following inductive result.
Assume the SM inequality (\ref{eq:strongmono}) holds for arbitrary pure states of $m<n$ qubits, then for the $n$-qubit states  $\ket{\Phi^n_{\alpha,\beta,\gamma}}$ one has:
${\tau}^{(1)} =  \frac{4}{n^2} \big[{n^2 |\alpha|^2 |\gamma|^2+(n-1)|\beta|^2(|\beta|^2+n|\gamma|^2)}\big]$, ${\tau}^{(2)} \leq \frac{4 |\beta|^4}{n^2}$, ${\tau}^{(n-1)} \leq \frac{4}{n}|\beta|^2 |\gamma|^2$,  ${\tau}^{(m)}=0$ for $2<m<n-1$.
Substituting these into Eq.~(\ref{eq:taunpure}), one finds: $\tau^{(n)}_{q_1|\ldots|q_n}(\ket{\Phi^n_{\alpha,\beta,\gamma}}) \geq 4 |\alpha|^2 |\gamma|^2 \geq 0$, which proves the SM inequality (\ref{eq:strongmono}) for the $n$-qubit states of Eq.~(\ref{eq:gengg}). As the SM  clearly holds for three-qubit states, this yields a complete analytical SM proof for generalized GHZ/$W$ superpositions $\ket{\Phi^4_{\alpha,\beta,\gamma}}$ of $n=4$ qubits, which embody archetypical representatives of genuine multiparticle entanglement.

\begin{table*}[bt]
\begin{tabular}{ll}
\hline \hline
Normal-form states $\ket{G^x}$ (unnormalized) & Bounds to the reduced three-tangles $\tau^{(3)}$ \\
\hline
$\begin{array}{l}\ket{G^1_{abcd}}=\frac{a+d}{2}(\ket{0000}+\ket{1111})+\frac{a-d}{2}(\ket{0011}+\ket{1100})\\
\qquad \quad +\frac{b+c}{2}(\ket{0101}+\ket{1010}) + \frac{b-c}{2}(\ket{0110}+\ket{1001})\end{array}$ & $\tau^{(3)}_{q_i|q_j|q_k}=0$  \\ 
$\begin{array}{l}\ket{G^2_{abc}}=\frac{a+b}{2}(\ket{0000}+\ket{1111})+\frac{a-b}{2}(\ket{0011}+\ket{1100}) \\
\qquad \quad + c (\ket{0101}+\ket{1010}) + \ket{0110}\end{array}$  &
$\tau_{q_i|q_j|q_k}^{(3)} \leq \frac{4 \left| c\right|  \sqrt{\big(a^2-b^2\big) \big({a^*}^2-{b^*}^2\big)}}{\left(\left| a\right| ^2+\left| b\right| ^2+2 \left| c\right| ^2+1\right)^2}$\\
$\begin{array}{l}\ket{G^3_{ab}}=a(\ket{0000}+\ket{1111}) + b(\ket{0101}+\ket{1010}) + \ket{0110} + \ket{0011}\end{array}$ &
$\tau^{(3)}_{q_1|q_2|q_3}\!=\tau^{(3)}_{q_1|q_3|q_4}\!=0, \tau^{(3)}_{q_1|q_2|q_4}\! = \tau^{(3)}_{q_2|q_3|q_4}\! \leq\frac{4|a| |b|}{{(1+|a|^2+|b|^2)}^2}$ \\
$\begin{array}{l} \ket{G^4_{ab}}=a(\ket{0000}+\ket{1111}) + \frac{a+b}{2}(\ket{0101}+\ket{1010}) + \frac{a-b}{2}(\ket{0110}+\ket{1001}) \\
\qquad \quad+ \frac{i}{\sqrt{2}}(\ket{0001} + \ket{0010} + \ket{0111} + \ket{1011})\end{array}$
& $\tau^{(3)}_{q_i|q_j|q_k}  \leq \frac{2|a^2-b^2|}{(2+3|a|^2+|b|^2)^2}$\\
$\begin{array}{l}\ket{G^5_a}=  a(\ket{0000} + \ket{0101} + \ket{1010} + \ket{1111}) +i\ket{0001} + \ket{0110} - i\ket{1011}\end{array}$ &
$\tau^{(3)}_{q_1|q_2|q_3}\!=\tau^{(3)}_{q_1|q_3|q_4}\! \leq \frac{16 |a|^2}{(3+4|a|^2)^2},\  \tau^{(3)}_{q_1|q_2|q_4}\! = \tau^{(3)}_{q_2|q_3|q_4}\! \leq\frac{4}{(3+4|a|^2)^2}$ \\
$\begin{array}{l}\ket{G^6_a}= a(\ket{0000} + \ket{1111}) + \ket{0011} + \ket{0101} + \ket{0110}\end{array}$ &
$\tau^{(3)}_{q_1|q_j|q_k}\!=0,\ \tau^{(3)}_{q_2|q_3|q_4}\! \leq
 \left\{ \begin{array}{cc}
 \frac{|a| \left(|a|^3-4\right)^2}{\left(2 |a|^2+3\right)^2} & |a|<2^{2/3} \\
 0 & |a|\geq 2^{2/3} \\
\end{array} \right.$
\\
$\begin{array}{l}\ket{G^7}= \ket{0000} + \ket{0101} + \ket{1000} + \ket{1110}\end{array}$ & 
$\tau^{(3)}_{q_1|q_j|q_k}\! \leq \frac14,\  \tau^{(3)}_{q_2|q_3|q_4}\!=0$\\
$\begin{array}{l}\ket{G^8}= \ket{0000} + \ket{1011} + \ket{1101} + \ket{1110}\end{array}$ & $\tau^{(3)}_{q_1|q_j|q_k}\! \leq \frac14,\  \tau^{(3)}_{q_2|q_3|q_4}\!=0$\\
$\begin{array}{l}\ket{G^9}= \ket{0000} + \ket{0111}\end{array}$ & $\tau^{(3)}_{q_1|q_j|q_k}\! =0,\  \tau^{(3)}_{q_2|q_3|q_4}\!=1$\\
\hline \hline
\end{tabular}
\caption{\label{tabbathehutt} Normal-form representatives of the nine four-qubit SLOCC classes defined in \cite{verstraete4}, and upper bounds to the three-tangle of their marginal three-qubit partitions $q_i|q_j|q_k$; here  $a,b,c,d$ are complex parameters with nonnegative real part.}
\end{table*}

{\it Four-qubit strong monogamy: Toolkit.}---
Motivated by the above result, we now analyze arbitrary pure states $\ket{\psi}$ of four-qubit systems ($n=4$). Here, a preliminary numerical exploration reveals that the choice $\mu_m=1$ in (\ref{eq:strongmono}) is too strong to hold, as it leads to negative residual four-tangles on a small subset of states \footnote{The states with negative residuals in the case $\mu_m=1$ were all found within the SLOCC class $4$ according to the classification of \cite{verstraete4}.}. Therefore, we focus on testing the SM inequality for a successive level of the hierarchy, namely we set  $\mu_m := m/2$ ($m \geq 2$).
Sticking with $q_1$ as focus, and according to our adopted conventions, the SM inequality (\ref{eq:strongmono}) then specializes to (see Fig.~\ref{fig:monineq} for a graphical representation)
\begin{equation}
\label{eq:ckw_ineq_gen}
\tau^{(1)}_{q_1|(q_2 q_3 q_4)} \geq \tau^{(2)}_{q_1|q_2}\!\!+\tau^{(2)}_{q_1|q_3}\!\!+\tau^{(2)}_{q_1|q_4}+[\tau^{(3)}_{q_1|q_2|q_3}]^{\frac{3}{2}}\!+[\tau^{(3)}_{q_1|q_3|q_4}]^{\frac{3}{2}}\!
+[\tau^{(3)}_{q_1|q_2|q_4}]^{\frac{3}{2}},
\end{equation}
where we omitted the state $(\ket{\psi})$ for brevity.

All the quantities in (\ref{eq:ckw_ineq_gen}) are well defined. The bipartite terms $\tau^{(m)}$ with $m=1,2$ are all computable as described above, and the tripartite terms $\tau^{(3)}$ are to be evaluated on the reduced rank-$2$ mixed state $\rho_{ijk}$ of qubits $q_i$, $q_j$, and $q_k$, via the prescription in Eq.~(\ref{eq:taunmix}). Let us recall that the three-tangle of three-qubit pure states $\ket{\psi}$ admits the following closed expression \cite{monogamy},
\begin{eqnarray}\label{eq:tau3pure}
\nonumber
\tau^{(3)}_{q_1|q_2|q_3}(\ket{\psi})&=& 4\big\vert c_{000}^2\, c_{111}^2+c_{001}^2\, c_{110}^2+c_{010}^2\, c_{101}^2+c_{100}^2\, c_{011}^2\\
\nonumber
&&-2(c_{000}\, c_{111}\, c_{001}\, c_{110}+c_{000}\, c_{111}\, c_{010}\, c_{101}\\
&&+c_{000}\, c_{111}\, c_{100}\, c_{011}+c_{001}\, c_{110}\, c_{010}\, c_{101}\\
\nonumber
&&+c_{001}\, c_{110}\, c_{011}\, c_{100}+c_{100}\, c_{011}\, c_{010}\, c_{101})\\
\nonumber
&&+4(c_{000}\, c_{011}\, c_{101}\, c_{110}+c_{111}\, c_{100}\,  c_{010}\, c_{001})\big\vert,
\end{eqnarray}
where we have expanded the state $\ket{\psi}$ in the computational basis as $\ket{\psi}=\sum_{r,s,t =0}^{1} c_{rst}\ket{rst}$.
However, to date, there is no closed formula for the three-tangle of three-qubit mixed states. The minimization in Eq.~(\ref{eq:taunmix}) has been solved only for special families of states \cite{lohmayer,lohmayernjp,jens,jens2}, while a semi-analytic method to determine when $\tau^{(3)}_{q_i|q_j|q_k}$ vanishes is generally available for rank-$2$ states such as $\rho_{ijk}$ \cite{lohmayer}. We then resort to looking for tractable upper bounds to the tripartite terms \cite{lohmayer, osborne, steepest, rdl, jens2}, say $\tau^{(3) \mathrm{up}}_{q_i|q_j|q_k} \geq \tau^{(3)}_{q_i|q_j|q_k}$. A lower bound to the residual four-tangle of Eq.~(\ref{eq:taunpure}) is then, for a four-qubit state $\ket{\psi}$,
\begin{equation}\label{eq:tau4low}
\tau^{(4) \mathrm{low}}_{q_1|q_2|q_3|q_4} :=  \tau^{(1)}_{q_1|(q_2 q_3 q_4)} - \sum_{j=2}^4 \tau^{(2)}_{q_1|q_j} - \sum_{k>j = 2}^4 \big[\tau^{(3) \mathrm{up}}_{q_1|q_j|q_k}\big]^{\frac32}\,, \end{equation}
and the SM inequality may then be verified by proving that $\tau^{(4) \mathrm{low}}_{q_1|q_2|q_3|q_4} \geq 0$. We will exploit in particular the bound  recently introduced by Rodriques, Datta, and Love (RDL) \cite{rdl} in terms of the so-called best $W$-class approximation of three-qubit states $\rho_{ijk}$ \cite{notew}.  For each rank-$2$ three-qubit state $\rho_{ijk} \equiv \rho$, which can be written in its spectral decomposition as $\rho = \lambda  \ketbra{1}{1} + (1-\lambda) \ketbra{2}{2}$, one can construct an associated simplex ${\cal S}_0$ containing states with vanishing three-tangle, obtained as mixtures of (up to) four pure $W$-class states $\ket{Z_l}$ ($l=1,\ldots,4$) \cite{notew}. The latter take the form $\ket{Z_l} = (\ket{1} + z_l \ket{2})/\sqrt{1+|z_l|^2}$, where $z_l \in \C$ are the complex roots of the fourth-order equation $\tau^{(3)}(\ket{1} + z \ket{2})=0$, defined via Eq.~(\ref{eq:tau3pure}) \cite{lohmayer}. If the rank-$2$ state $\rho$ belongs to the simplex ${\cal S}_0$, then $\tau^{(3)}(\rho)=0$. More generally, one can bound the three-tangle from above as follows. Defining the uniform mixture $\pi = \frac14 \sum_{l=1}^4 \ketbra{Z_l}{Z_l}$, there exists a $\kappa >0 $ such that
$\ketbra{\phi}{\phi}:=\rho+\frac{\kappa}{\|\,\rho-\pi\,\|_1}(\rho-\pi)$
describes a pure three-qubit state \cite{rdl}, where $\| X \|_1=\mathrm{Tr}\sqrt{X^\dagger X}$ denotes the trace norm. One has then
\begin{equation}\label{eq:taurdl}
\tau^{(3)}(\rho)\leq \tau^{(3) \mathrm{up}}(\rho):=  \frac{\|\,\rho-\pi\,\|^2_1}{\|\,\ketbra{\phi}{\phi}-\pi\,\|^2_1} \tau^{(3)}{(\ket{\phi})},
\end{equation}
where $\tau^{(3)}(\ket{\phi})$ can be computed from Eq.~(\ref{eq:tau3pure}).

{\it Four-qubit strong monogamy: Results.}---
For four qubits, there are infinitely many inequivalent SLOCC classes \cite{jensreview} (unlike the case of three qubits \cite{dur}); however, a particularly insightful classification into nine groups was derived by Verstraete {\it et al.}  \cite{verstraete4}, who showed that, up to permutations of the four qubits,  any  pure state $\ket{\psi}$  can be obtained as
\begin{equation}\label{4slocc}
\ket{\psi} = (A_1 \otimes A_2 \otimes A_3 \otimes A_4)\, \ket{G^x}\,,
\end{equation}
where $\{A_k\} \in SL(2,\mathbb{C})$ are SLOCC operations with $\det(A_k)=1$, and each $\ket{G^x}$ denotes a normal-form family of states, representative of the corresponding $x^{\rm th}$ class, with $x=1,\ldots,9$, see Table~\ref{tabbathehutt} for their definition; only class-$1$ states are generic.

\begin{figure*}[t!]
  \centering
    \includegraphics[width=8.5cm]{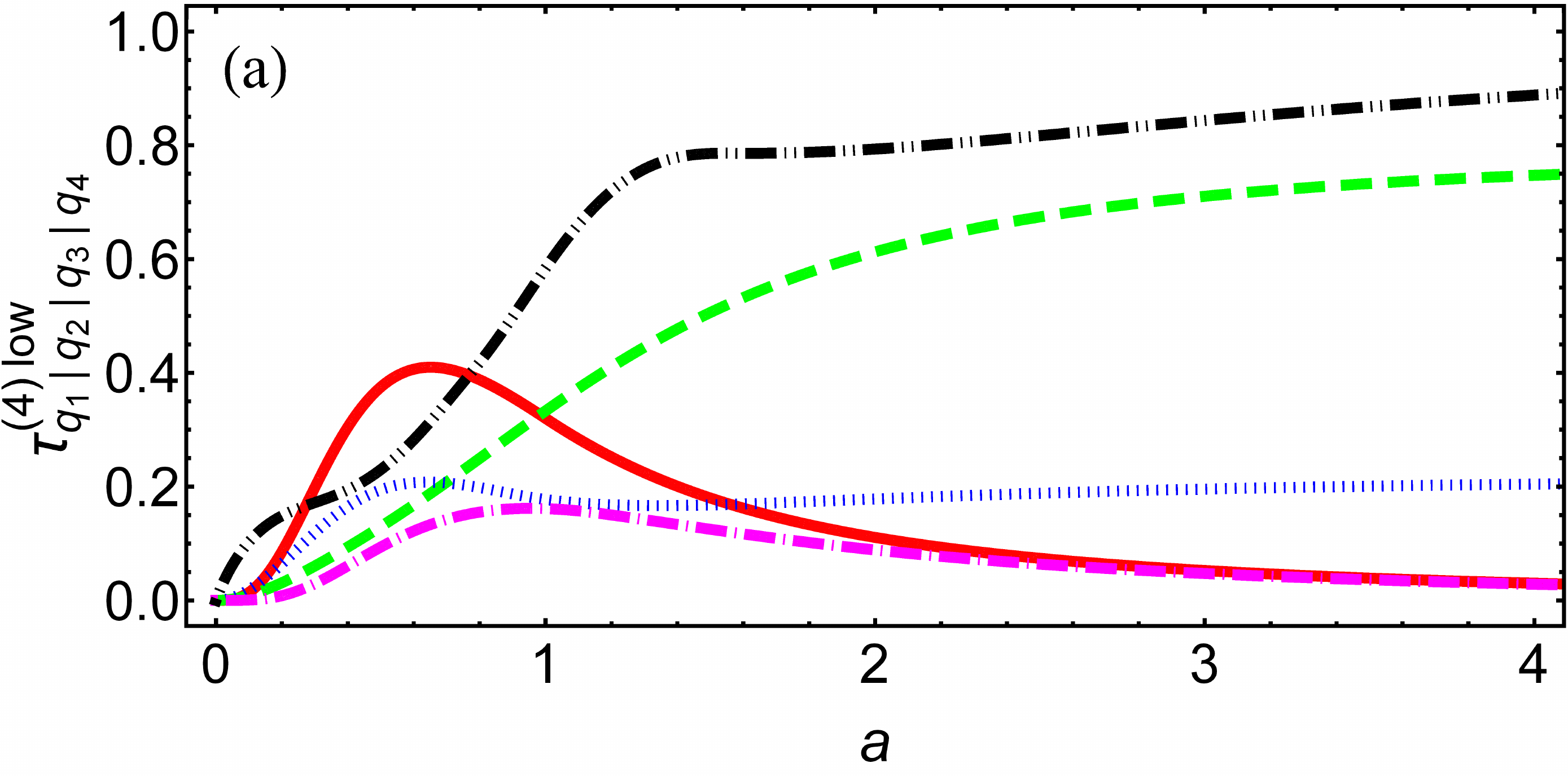} \hspace*{.5cm}
    \includegraphics[width=8.5cm]{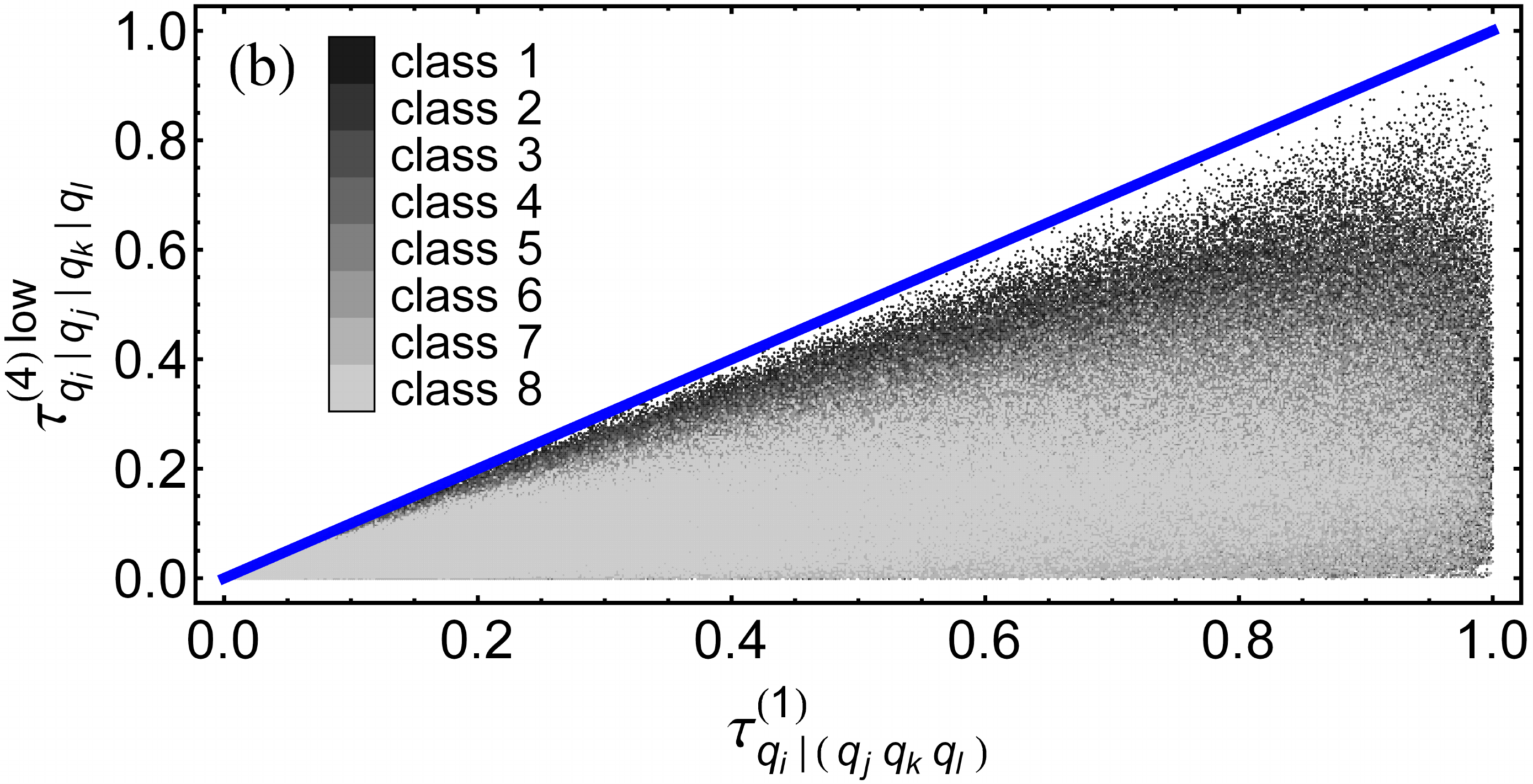} \\[-0.3cm]
 \caption{(Color online) (a) Lower bound to the residual four-tangle  $\tau^{(4)  \mathrm{low}}_{q_1|q_2|q_3|q_4}$ versus the  parameter $a$ (here assumed real) for the normal-form states: $\ket{G^2_{abc}}$ with $b=c=a$ (red solid line), $\ket{G^3_{ab}}$ with $b=a/4$ (green dashed line), $\ket{G^4_{ab}}$ with $b=a/2$ (blue dotted line), $\ket{G^5_a}$ (magenta dot-dashed line),  $\ket{G^6_a}$ (black dot-dot-dashed line). The residuals stay nonnegative for general choices of the parameters $a,b,c$.
(b) Lower bound to the residual four-tangle  $\tau^{(4)  \mathrm{low}}_{q_i|q_j|q_k|q_l}$ versus the  one-tangle $\tau^{(1)}_{q_i|(q_jq_kq_l)}$ for $8 \times 10^6$ random four-qubit pure states, with $4$ partitions tested per state. Each point is gray-scaled according to the SLOCC class of the state, from $90\%$ gray (darkest, class $1$) to $20\%$ gray (lightest, class $8$).  The solid line is saturated by GHZ states. All the data points are above the horizontal axis, verifying the SM inequality (\ref{eq:ckw_ineq_gen}).
 \label{fig:forza4}\vspace*{-.2cm}}
\end{figure*}

We verified the proposed SM inequality (\ref{eq:ckw_ineq_gen})  for the normal-form representatives   $\ket{G^x}$ of all the nine classes, by obtaining suitable analytic upper bounds to the $\tau^{(3)}$ terms in all the three-qubit marginal partitions, as presented in Table~\ref{tabbathehutt}.
Combining these bounds with the easily computable one-tangles $\tau^{(1)}_{(q_i|q_j q_k q_l)}$, and with the expressions of the reduced two-tangles $\tau^{(2)}_{q_i|q_j}$ (not reported here), we obtained lower bounds to the residual $\tau^{(4)}_{q_i|q_j|q_k|q_l}$ defined as in Eq.~(\ref{eq:tau4low}), which were found to be nonnegative for all the nine families of states. These are plotted in Fig.~\ref{fig:forza4}(a) for some typical instances of $\ket{G^x}$ with $x=2,\ldots,6$. The other cases are straightforward, in particular for $\ket{G^1}$ and $\ket{G^7}$ all the reduced three-tangles vanish, so the SM reduces to the conventional inequality (\ref{eq:gen_ckw}).

We complement this collection of analytical results  with a numerical exploration of arbitrary four-qubit states $\ket{\psi}$. Precisely, we generated them according to the prescription in Eq.~(\ref{4slocc}), by the application of random SLOCC operations on $\ket{G^x}$ states with randomized parameters (a Gaussian distribution was used to generate the matrix elements of SLOCC operations on each qubit, and a uniform distribution in a bounded interval was used to generate the complex parameters in the states $\ket{G^x}$). We tested $10^6$ states per class, and on each state we computed the lower bound  $\tau^{(4) \mathrm{low}}_{q_i|q_j|q_k|q_l}$ for all four independent permutations of $\{ijkl\}$, using the semi-analytical RDL method \cite{rdl} to bound the marginal three-tangles in all relevant three-qubit partitions via Eq.~(\ref{eq:taurdl}). Overall, this amounts to $3.2 \times 10^7$ tested data points across all the different classes   (class-$9$ states are excluded since for them   $q_1$ is separable from the rest, so the SM constraint reduces to the CKW one for   $q_2, q_3, q_4$ which needs no testing). As Fig.~\ref{fig:forza4}(b) shows, no negative values of $\tau^{(4) \mathrm{low}}$ were found, providing a strongly supportive evidence for the validity of the SM inequality (\ref{eq:ckw_ineq_gen}) on {\it arbitrary} four-qubit states.



{\it Conclusion.}---
We proposed and analyzed a novel class of monogamy inequalities for multiqubit entanglement, which extend and sharpen the existing ones \cite{monogamy,genmon}. We proved our SM relation on relevant families of states, and verified it numerically  on arbitrary pure states of four qubits spanning all the different  SLOCC classes of Ref.~\cite{verstraete4}.

This Letter opens an avenue for further investigation. First, a closed formula for the three-tangle of rank-$2$ states of three qubits \cite{lohmayer,lohmayernjp,jens} could facilitate a general analytical proof of inequality (\ref{eq:strongmono}) for $n=4$.
More generally, would other entanglement measures which satisfy conventional monogamy---such as the squashed entanglement \cite{squashed})---obey SM-type inequalities too, for arbitrary multipartite states of $n$ qudits?
The standard CKW-type monogamy \cite{monogamy,genmon}
inspired remarkable applications to quantum cryptography \cite{review_horodecki} and the characterization of quantum critical points in many-body systems \cite{review}. This work reveals more severe limitations on the sharing of multiple forms of entanglement, and is a starting point towards a quantification of those essential features of quantum correlations, which only emerge beyond the bipartite scenario. It will be fascinating to investigate the interplay between the SM trade-off and frustration phenomena in complex quantum systems \cite{frust1,frust2,frust3}.

{\it Acknowledgments.}---
We acknowledge discussions with M. Cianciaruso, N. Datta, C. Eltschka, P. Facchi, G. Florio, F. Illuminati, J. S. Kim, T. Osborne, S. Pascazio, M. Piani, J. Siewert, A. Winter, and W. K. Wootters. The authors would like to thank A.~Osterloh for bringing the error in the reduced three-tangle of class-2 states $\ket{G^2_{abc}}$ in the previously published version of this manuscript to their attention (see also Ref.~\cite{osterloh}).
This work has been supported by  the University of Nottingham, the University of Bari, the Italian National Group of Mathematical Physics (GNFM-INdAM), the Foundational Questions Institute (FQXi-RFP3-1317), and the Basic Science Research Program through the National Research Foundation of Korea funded by the Ministry of Education (NRF-2012R1A1A2003441).


\begin{thebibliography}{20}
\bibitem{notew}
By $W$-class states \cite{monogamy,dur,lohmayer} we refer to three-qubit pure states whose three-tangle $\tau^{(3)}$ as defined in Eq.~(\ref{eq:tau3pure}) vanishes.
\end{thebibliography}
\end{document}